\documentclass[%
 reprint,
superscriptaddress,
showpacs,
 amsmath,amssymb,
 aps,
]{revtex4-1}

\usepackage{graphicx}
\usepackage{color}
\usepackage{verbatim}
\usepackage{subfigure}

\definecolor{Red}{rgb}{1,0,0}

\begin{document}


\title{Fully autocompensating high-dimensional quantum cryptography  \\ by optical phase conjugation} 

\author{Jes\'us Li\~nares} 
\email{Corresponding author: suso.linares.beiras@usc.es}
\affiliation{Quantum Materials and Photonics Research Group, Optics Area, Department of Applied Physics, Faculty of Physics / Faculty of Optics and Optometry, University of Santiago de Compostela, Campus Vida s/n, E-15782, Santiago de Compostela, Galicia, Spain}
\author{Xes\'us Prieto-Blanco}
\affiliation{Quantum Materials and Photonics Research Group, Optics Area, Department of Applied Physics, Faculty of Physics / Faculty of Optics and Optometry, University of Santiago de Compostela, Campus Vida s/n, E-15782, Santiago de Compostela, Galicia, Spain}
\author{Daniel Balado}
\affiliation{Quantum Materials and Photonics Research Group, Optics Area, Department of Applied Physics, Faculty of Physics / Faculty of Optics and Optometry, University of Santiago de Compostela, Campus Vida s/n, E-15782, Santiago de Compostela, Galicia, Spain}
\author{Gabriel M. Carral}
\affiliation{Quantum Materials and Photonics Research Group, Optics Area, Department of Applied Physics, Faculty of Physics / Faculty of Optics and Optometry, University of Santiago de Compostela, Campus Vida s/n, E-15782, Santiago de Compostela, Galicia, Spain}

\begin{abstract} We present a bidirectional quantum communication system based on optical phase conjugation for achieving fully autocompensating high-dimensional quantum cryptography. We prove that random phase shifts  and couplings  among   $2N$ spatial and polarization optical modes described by  SU($2N$) transformations due  to  perturbations      are  autocompensated  after a  single round trip between Alice and Bob.   Bob can use a source of single photons or, alternatively,  coherent states and then Alice attenuates them up to a single photon level, 
 and thus non-perturbated 1-qudit states are generated for  high-dimensional QKD protocols, as  the BB84 one, of a higher security.  
   \end{abstract}

\maketitle 

{\em Introduction.$-$}Quantum cryptography is based on the properties of quantum mechanics to obtain secure quantum key distribution (QKD) by using different protocols. One of them is the seminal so-called BB84 protocol in which four states define a set of two mutually unbiased basis (MUBs). On the other hand, space division multiplexing has been proposed to further increase the data bandwidth in optical fiber communications \cite{Bai}, and  accordingly, high interest has arisen in new optical fibers such as few-mode fibers  and multicore fibers. Likewise,  optical satellite communications, and in general free space optical communications, based on spatial modes, such as those ones carrying orbital angular momentum, constitutes a promising communications technology \cite{Zou}.
\ \\
In parallel, the  interest in the implementation of quantum cryptography in both these new optical fibers and in free space has also remarkably increased in the last few years. The main reason is that  by using spatial optical modes a high-dimensional QKD (HD-QKD) can be implemented, which in turn improves  cryptographical security \cite{Ding}. Different optical  systems have been proposed to implement QKD cryptography in both optical fibers and free space; such systems can use different kind of modes, for instance, polarization modes in monomode optical fibers \cite{Muller} and free space \cite{Bedington}, collinear spatial modes in few-mode optical fibers \cite{Cozzolino} and free space   \cite{Jin} and spatial codirectional modes in multicore optical fibers \cite{Canas}. However, one of the most important drawbacks is that all guided and free space modes need to keep stable over long propagation distances along optical fibers or the atmosphere. Modes undergo  instability because  light, in its propagation in free space or along optical fibers, finds small spatial perturbations (imperfections) or slow temporal perturbations. This gives rise to random modal coupling (modal crosstalking)  which,  together with random intermodal phases,   causes instability of both modes and quantum states.   To overcome this drawback, specific (partial)  autocompensating techniques have been proposed in bidirectional quantum communication systems. For instance, polarization autocompensating quantum cryptography with 1-qubit  states excited in polarization modes \cite{Muller,Bethune}, or  more recently with 1-qudit states excited in spatial modes  acquiring random relative phases  \cite{Bal,Bal2}. However, to our knowledge, a fully autocompensating solution has not  been proposed.
\ \\
In this Letter we propose and prove a fully autocompensating quantum cryptography technique based on optical phase conjugation (OPC) and valid for both free space optical communication and optical fiber communications where spatial and polarization  modal couplings  are not negligible. In a most formal way, by using OPC we compensate for unwanted   effects in 1-qudit states caused by  an arbitrary number $q$ of unpredictable unitary transformations SU($2N$), where $2N$ is  the number of spatial modes with two polarizations. For the sake of expositional convenience we present  a detailed study for   multicore optical fibers (MCF) which can in turn  be formally  applied to both  few-mode optical fibers (FMFs) and free space communications.  Input multimode single photon states can be used, or alternatively, coherent states which are attenuated in their way back up to a single photon level (weak coherent states), so that  1-qudit states are produced. In this case, decoy states \cite{Hwang} have also to be generated for security purposes, as usual.  \begin{figure}[h]
\centering
\includegraphics[width=8.5cm,clip]{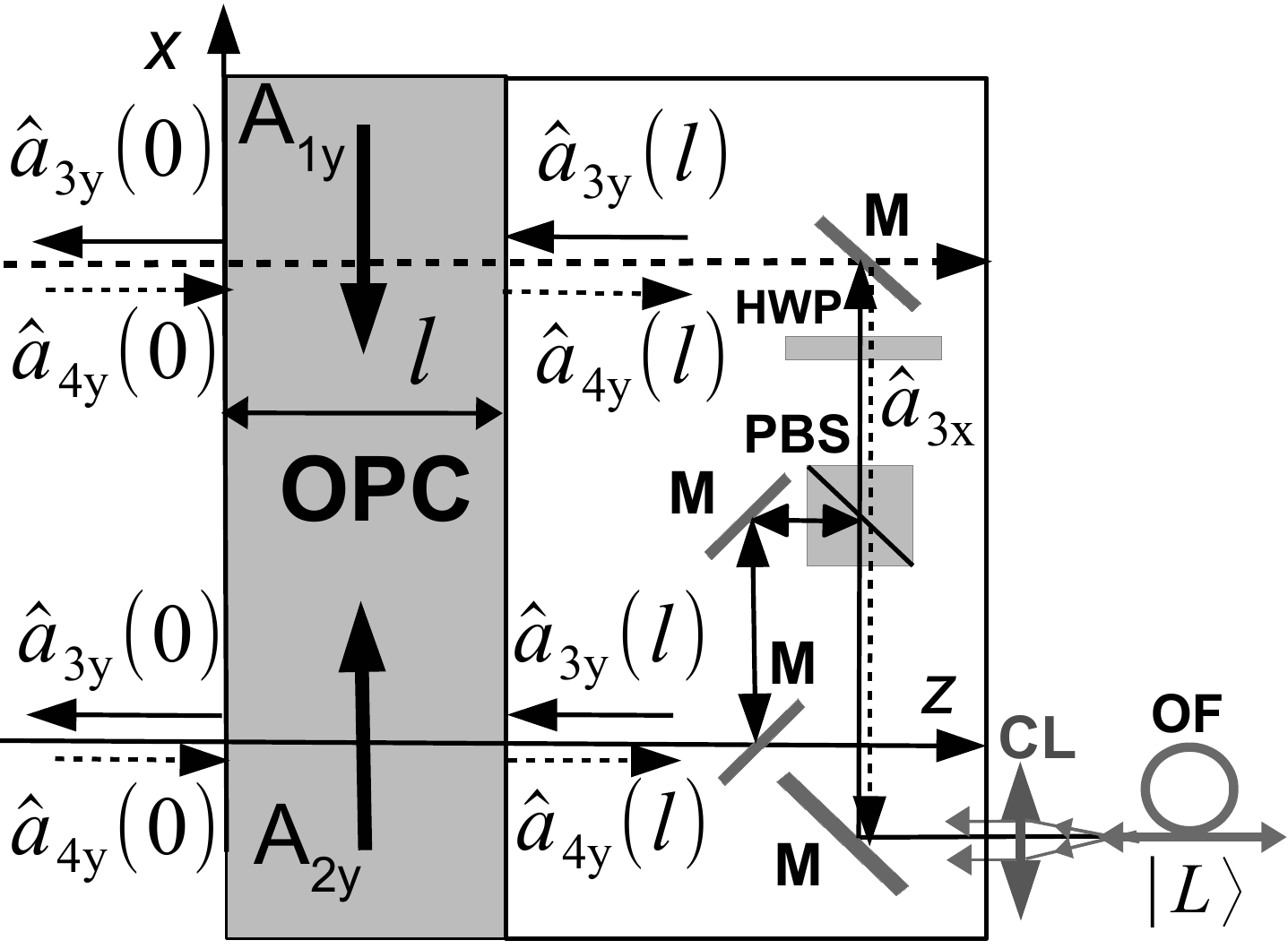}
\caption{Four-wave mixing  system with two pump waves ${\rm A}_{1y}$  and ${\rm A}_{2y}$ and an input state $\vert L\rangle$ emerging from  an optical fiber (OF) along $-z$. Optical fiber modes are collimated by a  lens (CL) and redirected by  mirrors (M).  PBS separates polarizations, and   vertical or $x$-polarization  becomes vertical or $y$-polarization  by using a HWP  rotated $\pi/4$.  Finally, the nonlinear medium implements an optical phase conjugation of the reflected modes $\hat{a}_{4y}$.}
\label{figPC}      
\end{figure}
\ \\
\textcolor{white}{sec}{\em Optical phase conjugator.$-$}Let us consider a  four-wave mixing (FWM) non-linear interaction system  as shown in Fig.~\ref{figPC}.  In general, an input spatial multimode optical quantum state $\vert L\rangle$  propagating along $-z$ direction and with frequency $\omega$  reaches the FWM system; such state emerges from an optical fiber (OF).  After the  OF  a collimating lens (CL) is inserted  for collimating the optical fiber modes ${\bf e}_{j}(x,y)$, $j$$=$$1,...,N$. This CL  is not strictly necessary but  helps to understand the physical process. After the collimating lens, a  polarizing beam-splitter (PBS) separates horizontal and vertical polarization modes ($x$ and $y$-modes). The linear $x$-polarization mode  is rotated $\pi/2$ by means of a HWP$_{\pi/4}$ (HWP  rotated $\pi/4$). Finally,  we have a third order non linear material  of length $l$ in which  it is considered that there are two strong  vertically polarized counter-propagating intense pump modes (strong coherent states) of frequency $\omega$ and   amplitudes ${\rm A}_{1y}$ and ${\rm A}_{2y}$.  A possible isotropic non linear material for implementing OPC can be  CS$_{2}$ \cite{Yariv}, which  would contribute for vertical polarization    with the term \smash{$\chi_{yyyy}^{(3)}$} of the  third order non linear tensor.  For the moment, let us consider that   $\vert L\rangle$  is a single mode state   excited in the incident  mode 3 coming from an OF and with an associated   optical field operator \smash{$\hat{\rm E}_{3y}$\,$\propto$\,$ \hat{a}_{3y}$}, with  $\hat{a}_{3y}$ an absorption operator. When $\vert L\rangle$   reaches the non linear material then a fourth mode (reflected mode 4) arises, with an  associated field operator \smash{$\hat{\rm E}_{4y}$\,$\propto$\,$\hat{a}_{4y}$}. We are interested on the quantum states propagating along $z$-direction in  modes 3 and 4 (idler and signal modes) after non linear interaction, that is, in spatial propagation  not in temporal evolution (Hamiltonian operator)  \cite{Ben-Aryeh}.  Therefore, for spatial nonlinear coupling  is convenient to use the Momentum  operator describing the quantum mode  interaction  \cite{Yariv, Ben-Aryeh, LinaresNJP}, 
\begin{equation}\label{Momentum}
\hat{M}_{I}=  \int \chi_{yyyy}^{(3)} {\rm E}_{1y}{\rm E}_{2y}\hat{\rm E}_{3y}\hat {\rm E}_{4y}\,dxdydt,
\end{equation}
where the intense pump waves 1 and 2 can be   treated clasically and modes 3 and 4 in a quantum way, that is, 
\begin{eqnarray}
{\rm E}_{({1y\atop 2y})}= {\rm A}_{({1y\atop 2y})}\,e^{\mp ik_{o}nx} \,e^{-i\omega t}+ c.c.,\\
\hat{\rm E}_{({3y\atop 4y})}= \sqrt{\hbar \omega}\,{\rm{e}}_{j}^{({3\atop 4})}(x,y)\,\hat{a}_{({3y\atop 4y})} \,e^{\mp ik_{o}nz} \,e^{-i\omega t}+ h.c.,
\end{eqnarray}
 where ${\rm e}_{j}$ is a $j$th normalized spatial mode of the OF with $x$ or $y$-polarization. As  these modes   are collimated by the CL then  \smash{${\rm{e}}_{j}^{(3)}$\,$\approx$\,${\rm{e}}_{j}^{(4)}$}. By inserting the above pump waves and operators into Eq.~(\ref{Momentum}) and performing the temporal integrations the following operator is found
\begin{equation}\label{Moper}
\hat{M}_{I}=\hbar \chi_{eff} {\rm A}_{1y} {\rm A}_{2y} \hat{a}_{3y}^{\dag}\hat{a}_{4y}^{\dag} +h.c.,
\end{equation}
where $\chi_{{\text {\em eff}}}$ is an effective non-linear susceptibility  which  groups  together all physical constants.  From this operator the spatial Heisenberg equations  \cite{Ben-Aryeh} can be obtained, \begin{equation}\label{Heis}
-i\hbar\frac{\partial \hat{a}_{my}}{\partial z}=[\hat{a}_{my},\hat{M}_{I}],
\end{equation}
 where $m$\,$=$\,$3,4$.   Let us  denote the input operators as  $\hat{a}_{3y}(l)$\,$\equiv$\,$\hat{a}_{o3}$ and  $\hat{a}_{4y}(0)$\,$\equiv$\,$\hat{a}_{o4}$, and  the output operators as $\hat{a}_{3y}(0)$\,$\equiv$\,$\hat{a}_{3}$, associated to the optical mode transmitted  along system, and $\hat{a}_{4y}(l)$\,$\equiv $\,$\hat{a}_{4}$, associated to the reflected optical mode (see Fig.~\ref{figPC}). As mentioned,  there are  two pump waves with very  large amplitudes {A}$_{1y}$ and {A}$_{2y}$ and initial phases equal to zero, then the non-linear interaction strength is given by  a  coupling coefficient  $\kappa$\,$=$\,$\chi_{{\text {\em eff}}}\,\vert\rm{A}_{1y}\vert\vert\rm{A}_{2y}\vert$,  therefore the coupling is parametrically governed by $\vert\rm{A}_{1y}\vert\vert\rm{A}_{2y}\vert$, that is, the efficiency of the process is governed by pumping.  It is easy to check  that the solutions of the spatial Heisenberg equations obtained by inserting  Eq.~({\ref{Moper})  into  Eq.~(\ref{Heis}) provide  the well-known  operator transformations \cite{Yariv, Mandel}
 \begin{equation}\label{ecA}
\hat{a}_{3y}(l)\equiv\hat{a}_{o3}=\sec(\kappa l)\,\hat{a}_{3}+i\tan(\kappa l)\,\hat{a}_{4}^{\dag}
\end{equation}
\vspace{-0,7cm}
\begin{equation}
\hat{a}_{4y}(0)\equiv\hat{a}_{o4}=\sec(\kappa l)\,\hat{a}_{4}+i\tan(\kappa l)\,\hat{a}_{3}^{\dag}.
\end{equation}
Next, let us consider a  coherent state $\vert L\rangle=\vert \alpha_{3} 0_{4}\rangle$ excited in an optical fiber mode.  By taking into account the complex displacement operator and using Eq.~(\ref{ecA}) the output state can be rewritten as follows
\begin{equation}
\vert L\rangle=
e^{\alpha\hat{a}_{o3}^{\dag}-\alpha^{\star}\hat{a}_{o3}^{}}\,\vert 00\rangle\rightarrow \vert L_{c}\rangle=\vert s \alpha\rangle \,\vert \text{-}it\alpha^{\star}\rangle,
\end{equation}
where $s$$=$$\sec(\kappa l)$ and  $t$$=$$\tan(\kappa l)$. Note that  the reflected coherent state has been conjugated, that is, the FWM is an OPC; besides,  quantum-mechanically the OPC means that \smash{$\hat{a}_{o3}$\,$\propto$\,$ \hat{a}_{4}^{\dag}$}. Let us recall, however, that we are interested in multimode coherent states, that is, $\vert L\rangle$\,$=$\,$\vert \alpha_{1} ... \alpha_{N}\rangle$ excited in $N$ optical modes,  therefore, the OPC produces the multimode  coherent state  
$\vert L_{c}\rangle=(\vert s \alpha_{1}\rangle \,\vert \text{-}it\alpha_{1}^{\star}\rangle)\,...\,(\vert s \alpha_{N}\rangle \,\vert \text{-}it\alpha_{N}^{\star}\rangle)$.
Alternatively, we could have a single photon source, therefore we can generate an input  1-qudit state $\vert L_{o}\rangle= \sum_{o3j}c_{j}\vert 1_{o3j}\rangle$, with, $j$\,$=$\,$1,...N$, where the subindex $o3j$ indicates that the $N$ modes are incident on the OPC. By taking into account that $\vert 1_ {o3j}\rangle=(\hat{a}_{o3j}^{\dag}+\hat{a}_{o3j})\vert 0\rangle$, we obtain the output quantum state $\vert L\rangle= s\sum_{3j}c_{j}\vert 1_{3j}\rangle+it \sum_{4j}c_{j}\vert 1_{4j}\rangle$. \ \\
\textcolor{white}{sec}{\em Autocompensation with codirectional modes.$-$}First of all and for the sake of expositional convenience, we provisionally assume  that polarization is maintained  under propagation (we will consider spatial and  polarization couplings later, although in atmosphere and special optical fibers polarization can be maintained),  therefore we only  consider coupling among  $N$ spatial modes with the same linear polarization. In particular,  let us consider codirectional modes of a MCF with  $N$  modes (cores)   whose propagation constants are $\beta_{oi}, \, i$\,$=$\,$1,...,N$ and with associated absorption operators $\hat{a}_{i}$.   We must stress that the results that we are going to obtain are also valid for collinear modes of  FMF or a free space optical modes.    The quantum state reaching Bob system from the Alice system will be an unpredictable quantum state, and as a consequence,  modal coupling  prevents us to implement any QKD protocol. Next, we show how to overcome this drawback by OPC. Let us consider a perturbation  $P_{s}(x,y)$ that induces modal coupling and  can be considered as $z$-invariant  along a distance $s$, then the   
 spatial Heisenberg equation describing  modal coupling among $N$ optical field operator modes $\hat{E}_{i}$\,$\propto$\,$ \hat{a}_{i}$ can be written as follows \cite{LinaresNJP} 
 \begin{equation}\label{eqsystem}
-i\hbar\frac{\partial \hat{a}_{i}}{\partial z}=\hbar \{\beta_{i}\sum_{j=1}^{N}\delta_{ij}\hat{a}_{j}+\sum_{j\neq i}^{N}\kappa_{ij}\hat{a}_{j}\}\equiv \hbar\sum_{j=1}^{N}C_{ij}\hat{a}_{j},
\end{equation}
here $\beta_ {i}$\,$=$\,$\beta_{oi}$$+$$\kappa_{ii}$ are the perturbed propagation constants  due to (random) modal selfcoupling ($\kappa_{ii}$),
 and $\kappa_{ij}$ are (random) modal coupling coefficients due to cross coupling.  From a most fundamental point of view, an arbitrary  coupling coefficient of spatial  modes $i,j$  is given by $\kappa_{ij}$\,$=$\,$ \int {\rm e}_{i}(x,y) P_{s}(x,y) {\rm e}_{j}(x,y)dxdy$,  
where ${\rm e}_{(i,j)}(x,y)$ are the mode amplitudes. 
Note that  $\kappa_{ij}$\,$=$\,$\kappa_{ji}$ and therefore   $[C_{ij}]$\,$\equiv $\,$C$ is a symmetric matrix.    Therefore, by using the algebraic properties of symmetric matrices, the formal matrix solution $[S_{ij}]$\,$\equiv $\,$S$\,$=$\,$\exp\{iCz\}$ of differential equation in Eq.~(\ref{eqsystem}) is  a complex symmetric matrix, that is, $S_{ij}$\,$=$\,$S_{ji}$. On the other hand, modal coupling is an unitary transformation, therefore \smash{$[S_{ij}]^{-1}$\,$=$\,$[S_{ji}]^{\star}$\,$=$\,$[S_{ij}]^{\star}$}. In general, we will have an arbitrary number $q$ of perturbations, accordingly, the total effect along $z$ direction  from a  system B (Bob) to system A (Alice) can be  represented by the total matrix $M$\,$=$\,$S_{1}\cdots S_{q}$. Then, if we have an OPC at A, the matrix $M$ becomes $M^{\star}$\,$=$\,$S_{1}^{\star}\cdots S_{q}^{\star}$. Now, the quantum state is propagated back to system B, therefore  we have to use a  reflected coordinate system which is defined, without loss of generality, by $(-x)y(-z)$ with respect to the incident coordinate system $xyz$. The coupling coefficients  $\kappa_{ij}$ are invariant under the transformation $x$\,$\rightarrow$\,$ -$\,$x$ and    consequently the matrices $C$ are also invariant. Once the light  has travelled the path back and forth, the total coupling matrix is $M^{t}M^{\star}$\,$=$\,$\mathbb{I}$, with super index {\small{$t$}} indicating transpose. Then, the unpredictable modal coupling has been removed.  In short, if Bob launches a state $\vert L\rangle$ undergoing modal coupling along an OF or in the atmosphere,  the state after the OPC and  traveling its way back is  $\vert L_{c}\rangle=(\vert s \alpha_{1}\rangle \,\vert \text{-}it\alpha_{1}^{\star}\rangle)\,...\,(\vert s \alpha_{N}\rangle \,\vert \text{-}it\alpha_{N}^{\star}\rangle)$, that is, we recover the initial state except phases $\pi$ and conjugations.  \ \\
\textcolor{white}{sec}{\em Autocompensation with spatial and polarization  modes.$-$}As commented, we also have to  remove unpredictable polarization modal coupling together with  the  above spatial modal coupling. First of all we characterize the matrix transformation produced by an arbitrary coupling between linearly polarized spatial  modes (e.g., LP modes). Since there are $2N$ modes we consider the new subindices $i,j $\,$=$\,$\{1H,1V, 2H,2V, ..., NH,NV\}$, with $H$\,$\equiv $\,$x$, $V$\,$\equiv$\,$ y$. As in the above (scalar) case, the coupling  matrix $[C_{ij}]$\,$\equiv$\,$ C$ is also a complex symmetric matrix. However, when considering back propagation,  this matrix gets modified because, as commented, the incident coordinate system $xyz$  becomes $(-x)y(-z)$ under reflection. Indeed, an arbitrary  coupling coefficient of spatial  modes $m,n$ with different polarization is given by $\kappa_{mHnV}$\,$=$\,$ \int {\rm e}_{mH}(x,y) P_{v}(x,y) {\rm e}_{nV}(x,y)dxdy$, 
where $P_{v} (x,y)$ is  an arbitrary perturbation producing polarization (vector) modal coupling. Obviously, under reflection (back path) we have ${\rm e}_{nH}(x,y)$$\rightarrow$$-$${\rm e}_{nH}(x,y)$, then $\kappa_{ij}$$\equiv$$\kappa_{mHnV}$\,$\rightarrow$\,$-\kappa_{mHnV}$$\equiv $$-$$\kappa_{ij}$. Note that for the same polarization the coupling coefficient is positive (or zero) under reflection.  Therefore, the coupling matrix $[B_{ij}]$\,$\equiv$\,$ B$ under reflection can be written as follows
 \begin{equation}
B= (I_{N}\otimes\sigma_{z}) \,C \,(I_{N}\otimes\sigma_{z}) \equiv D\,C\, D,
\end{equation}
with $I_{N}$ the $N$-dimensional identity matrix,  $\otimes$ the  tensor product and $\sigma_{z}$  the third Pauli matrix. Next,  by taking into account that $DD$\,$=$\,$(I_{N}\otimes\sigma_{z})(I_{N}\otimes\sigma_{z})$\,$=$\,$I_{2N}$ ($I_{2N}$  the identity matrix $2N{\rm x}2N$), it is easy to check  that the transformation matrix produced by the perturbation $P_{v}(x,y)$, that is, the formal matrix solution $[R_{ij}]$\,$\equiv$\,$ R$\,$=$\,$\exp\{iBz\}$ can be written, by using the Taylor expansion of an exponential function, as follows
 \begin{equation}
R=(I_{N}\otimes\sigma_{z}) \,S \,(I_{N}\otimes\sigma_{z})\equiv D\,S\, D.
\end{equation}
  where $S$$=$$\exp\{iCz\}$. Note that matrix $R$ (transformation of the absorption operators $\hat{a}_{jH},\hat{a}_{jV}$) is also symmetric. On the other hand, it is also easy to check that the HWP$_{\pi/4}$ introduces  a phase $\pi$  between $H$-mode and $V$-mode of every spatial mode  in its  way back. Therefore, after OPC system, when polarization modes are recombined in the PBS (see  Fig.~\ref{figPC}),  the matrix  $D$\,$=$\,$I_{N}\otimes\sigma_{z}$ is implemented.   Consequently,  by  considering the general case of $q$ random couplings,  we obtain, after  the path back to Bob, the total matrix
\begin{equation}
M_{T}=R_{q}\cdots R_{1} D  S_{1}^{\star}\cdots S_{q}^{\star}=D,
\end{equation}
where we have taken into account the following relationships $R_{k}$\,$=$\,$ DS_{k}D$, $DD$\,$=$\,$I_{2N}$ and $S_{k}^{}S_{k}^{\star}$\,$=$\,$I_{2N}$, $k$\,$=$\,$1,...q$. In short, symmetric spatial perturbations together with polarization perturbations have been removed. 
 \ \\
\textcolor{white}{sec}{\em General {\rm SU(}$2N${\rm )} autocompensation.$-$}The above results have made clear autocompensation   of symmetric unitary coupling transformations. Now, we    generalize the above results for  non symmetric unitary coupling transformations, for example, rotations due to optical activity ($N$$=$$1$),   $2N{\rm x}2N$ abstract rotations and so on, that is, SU($2N$)  perturbations.  In order to prove this assertion we must take into account that all unitary transformation SU(2N) can be factorized as a ordered product of SU($2$) transformations of subspaces $i,j$ \cite{teoremaSU(N),Soto}. Thus, by proving autocompensation by OPC of an arbitrary  SU($2$) transformation the case SU($2N$) is also proven. A general SU(2) transformation $S$ can be represented by 
\begin{equation}\label{ZXZ}
S=\begin{pmatrix}\cos\theta& i\sin\theta e^{-i\delta}\\  i\sin\theta e^{i\delta}&\cos\theta\end{pmatrix}\equiv Z({\delta}) X({\theta}) Z({-\delta})
\end{equation}
with $Z(\pm\delta)$$=$$diag(1,e^{\pm i\delta})$ the matrix of a phase retarder $\pm\delta$ generated by the the Pauli's matrix $\sigma_{z}$, and $X({\theta})$  a matrix whose generator is the Pauli's matrix $\sigma_{x}$, with elements $X_{11}({\theta})$$=$$X_{22}({\theta})$$=$$\cos\theta$, and $X_{12}({\theta})$$=$$X_{21}({\theta})$$=$$i\sin\theta$. By considering that we are in a   polarization two-mode subspace, it is easy to check that the matrix by reflection  is characterized by the changes $\delta,\theta $\,$\rightarrow$\,$ -\delta,-\theta$, then $T$\,$=$\,$D_{2}S^{T}D_{2}$,  where $D_{2}$\,$=$\,$\sigma_{z}$. Therefore, after OPC and by taking into account the action of the HWP$_{\pi/4}$ ($D_{2}$) we obtain the following result
\begin{equation}
T S^{\star}=D_{2}S^{T}D_{2}\,D_{2}\,S^{\star}=D_{2}S^{T}\,S^{\star}= D_{2},
\end{equation}
and then autocompensation is again achieved. 
Rigorously, a general SU(2) matrix  requires to consider an additional matrix corresponding to  a  $\alpha$-phase retarder, that is, $Z(\alpha)S$; however, a retarder is also autocompensated, as proven above. We must also stress  that   similar results are found for   spatial  two-mode subspace. Likewise,    topological phases due to helical paths, torsions and so on of  an OF can be also autocompensated because  such phases  also correspond to unitary transformations (rotations and so on) \cite{Berry}.  In short, a multimode coherent state $\vert L\rangle$\,$=$\,$\vert \alpha_{1H} \alpha_{1V} ... \alpha_{NH}\alpha_{NV}\rangle$ coming from Bob, becomes a predictable reflected multimode coherent state $\vert L_{c}\rangle$\,$=$\,$\vert  $${-}$$it\alpha_{1H}\, it\alpha_{1V} ... $${-}$$it\alpha_{NH}\,it\alpha_{NV}\rangle$, therefore, we have proved that OPC has canceled a number $q$ of unpredictable perturbations represented by  SU(2$N$) transformations.
\ \\
\textcolor{white}{sec}{\em Optical fiber  setup for HD-QKD.$-$}By taking into account the  results obtained   we can implement an  autocompensating optical system for  HD-QKD  BB84 quantum cryptography in optical fibers.  Such a system is shown in the sketch of Fig.~\ref{figsystem}. Its optical configuration  is as follows: 
the first device is a coherent states generator (CSG) located in Bob system, which emits  coherent states $\vert L\rangle$\,$=$\,$\vert \alpha_{1H}  \alpha_{1V} ... \alpha_{NH} \alpha_{NV} \rangle$ to be excited in $2N$ optical modes of a MCF (or a FMF fiber, or  free space), with $\alpha_{1H}$\,$=$\,$...$\,$=$\,$\alpha_{NV}$, for  generating quantum states belonging to MUBs. Next, optical fiber delayers (OFD$_{1}$) produce modal delays $\tau_{j} \,(j$\,$=$\,$1, ...,2N)$, that is, we have a multimode coherent state formed by the tensor product of delayed single mode coherent states.  These delays will allow to Alice introduce phases in each spatial mode. Afterwards, a set of optical circulators  SOC   launches the state towards Alice system. 
 Besides, a multiplexing/demultiplexing ({\sc mux/demux})  device is also  needed if  collinear propagation is required as in a  FMF or even in free space.   Different spatial multiplexing  devices can be implemented according to the kind of collinear modes \cite{LinaresOL,OAMsorter}. Note that If we use a  MCF then  the  {\sc mux/demux}  device is not required (codirectional modes).
\begin{figure}[h]

\centering
\includegraphics[width=8.5cm,clip]{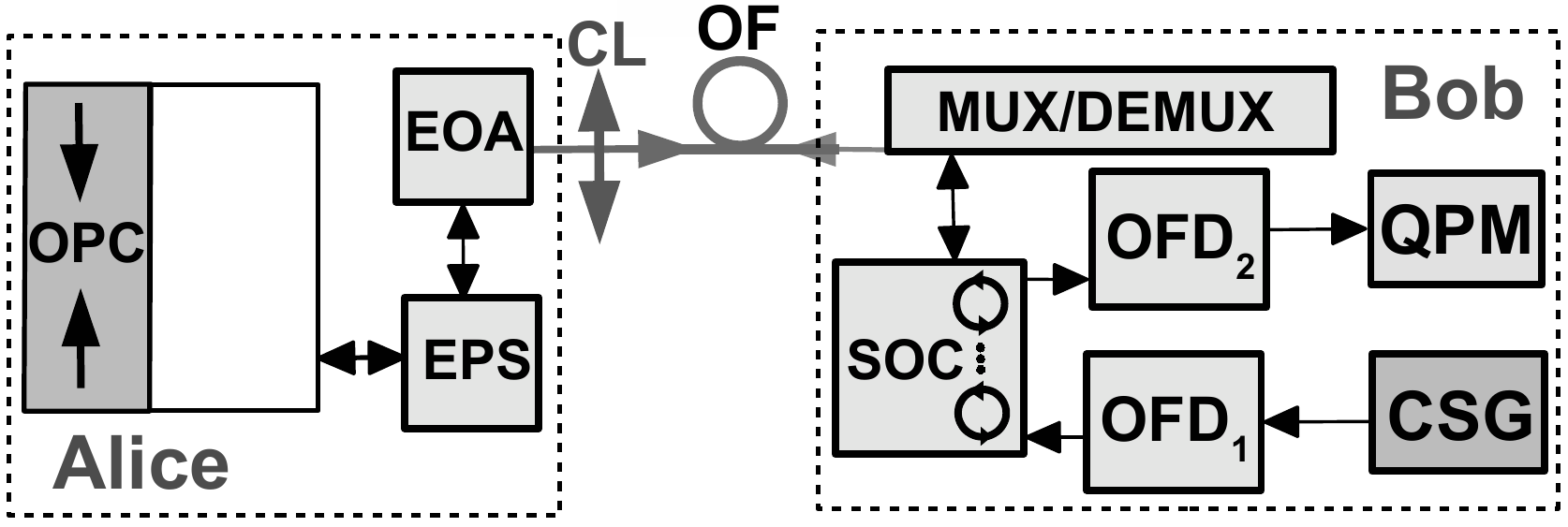}
\caption{Basic optical fiber setup for  autocompensating HD-QKD  by OPC (see text under {\em Optical fiber setup for HD-QKD}  for description).}
\label{figsystem}       
\end{figure}
After propagation along the OF  each $j$th delayed single mode coherent state (excited in each core of a MCF) becomes 
 a multimode coherent state due to  modal coupling. Such an state reaches the OPC device described in detail in Fig.~\ref{figPC} and explained above. Note that now  we have placed an electro-optic  phase shifter (EPS) and an electro-optic attenuator (EOA) \cite{Bal}  in off-position between the OPC and the CL-OF (see Fig.~\ref{figsystem}).  We must stress that  after the OPC the  reflected state has to be coupled again to the OF, however, the OPC  implements by itself this transverse  modal coupling to the OF (analogous  to the well-known image restoration by OPC \cite{Yariv,Mandel}), although different systems can be used to optimize this coupling, for example, as in our case,  by means of a CL.   Next, the reflected state in the OPC device goes through the EPS which introduces global phases $\theta_{j}$ on the mentioned multimode coherent states, and the EOA attenuates the state up to single photon level (note that at Bob system  each of these multimodes states will become again a $j$th delayed single mode coherent state due to OPC).  The purpose of the EOA is not only to attenuate  but also to increase the security of the system, i.e., the attenuation of the EOA can be controlled, enabling the  production of different attenuated pulses: signal and decoy states  against different attacks of an eavesdropper Eve, as in the photon-number-splitting attack, although we must stress that modal coupling is also a defense against attacks in line. Therefore,  Alice system  generates  1-qudit  states which propagate along the OF up to the Bob system  and thus the modal coupling and relative phases are fully removed. Next, the set of optical circulators SOC sends the 1-qudit state to the OFD$_{2}$ device which  cancels the delays $\tau_{j}$ between states $\vert 1_{j}\rangle$ of the 1-qudit. In short,  the quantum state generated is 
 \begin{equation}\label{Lcfinal}
 \vert L_{c}\rangle \approx  \frac{-i}{\sqrt{2N}}\,\big\{\sum_{j=1}^{N}\,e^{i\theta_{jH}}\vert 1_{jH}\rangle -      \sum_{j=1}^{N} e^{i\theta_{jV}}\vert 1_{jV}\rangle\big\}.
\end{equation}
These states  allow to implement pairs of $2N$-dimensional MUBs \cite{Ding} for QKD.  Finally,  the state reaches a  quantum projective measurer (QPM). Since both  QMP and CSG    are  relevant devices  it is worth showing a possible physical implementation of them.  A  CSG can be easily  made  by integrated devices with concatenated  directional couplers  2{\rm x}2, that is,  each output of a coupler is  connected to other coupler and so on \cite{Ding,Bal}. Such couplers can be represented by  matrices $X(\theta)$ such as the one presented in Eq.(\ref{ZXZ}) where $\theta$$=$$\kappa  d$ with $\kappa$ a linear coupling coefficient and $d$ the coupling distance.  We  start from  a single-mode coherent state $\vert L_{sm}\rangle$\,$=$\,$\vert\beta\rangle$, and by modal coupling with concatenated directional couplers $X(\pi/4)$ a multimode state $\vert L'_{sm}\rangle=\vert \alpha\rangle  \vert i \alpha\rangle\vert$$-$$ \alpha\rangle \vert i \alpha\rangle ...\vert $$-$$ \alpha\rangle \vert $$-$$i \alpha\rangle \vert $$-$$ \alpha\rangle \vert i \alpha\rangle$ is obtained, where $\alpha$\,$=$\,$2^{-N/2}\beta$.  
The relative phases $\{\pm \pi/2,\pi\}$ can easily be cancelled by using the EPS of the Alice system when the  proper phases $\theta_{1H}, ...\theta_{NV}$ are introduced, then $\vert L\rangle=\vert \alpha\rangle...\vert\alpha\rangle$. 
As to the QPM, a passive integrated quantum projective measurer that randomly selects bases of  dimension $N$$=$$2^{m}$ in MCFs has  been recenlty proposed   by using \smash{$X(\pi/4)$} and \smash{$X(\pi/2)$} couplers and phase shifters \smash{$Z(\delta)$} \cite{Bal}. Obviously, if we had used single photon sources both the attenuation and decoy states would not be  required.  Finally, we must indicate that bidirectional QKD systems are subject to new lateral attacks like the phase-remapping (PR)  one. Likewise,  recent   security analysis  considering such an  attack have been made  \cite{Bal,Bal2}, and it was shown that PR  can not reduce the QBER bellow the minimum   in which  a secret key rate is guaranteed under a cloning attack in a normal one-way system. Therefore, bidirectionality does not necessarily make the system more vulnerable.
\ \\
\textcolor{white}{sec}{\em Free space optical communications.$-$}As commented, free space optical communications can be considered as a problem of $2N$ collinear modes. As in the case of  FMFs, the generation and measurement of quantum states have to be made by a {\sc mux/demux} process.  A detailed study of this case would incorporate mode diffraction, however in most of cases it can be reduced thanks to the high directionality of lasers, or it can simply  be ignored  since OPC also compensates diffraction (note that a diverging wave incident on an OPC becomes a converging wave).
\ \\
\textcolor{white}{sec}{\em Summary.$-$}
We have presented a fully autocompensating technique based on optical phase conjugation for high-dimensional quantum cryptography in optical fibers and free space. A single round trip allows to auto-compensate the undesired modal coupling and random  phase shifts among spatial and polarization modes, and thus HD-QKD  protocols such as the BB84  can be implemented. 
\ \\
\textcolor{white}{sec}Authors wish to acknowledge the financial support of this work by Xunta de Galicia, Consellería de Educación, Universidades e FP, with a grant Consolidation-GRC Ref.-ED431C2018/11, grant  Strategic Grouping of Materials (AeMAT)  Ref.-ED431E 2018/08, and a predoctoral grant (D. Balado, 2017), co-financied with the European Social Fund.



\begin{thebibliography}{15}
\bibitem{Bai} N. Bai, E. Ip, Y. Huang, E. Mateo, F. Yaman, M. Li, S. Bickham, S.
Ten, J. Li\~nares, C. Montero, V. Moreno, X. Prieto, V. Tse, K. M. Chung, A. P. T. Lau, H. Tam, C. Lu, Y. Luo, G. Peng, G. Li, and T. Wang, Mode-division multiplexed transmission with inline few-mode fiber amplifier, Opt. Express {\bf 20}, 2668 (2012)
\bibitem{Zou} L. Zou, X. Gu and L. Wang, High-dimensional free-space optical communications based on orbital angular momentum coding, Opt. Commun. {\bf 410} 333 (2018)
\bibitem{Ding} Y. Ding, D. Bacco, K. Dalgaard, X. Cai, X. Zhou, K. Rottwitt, and L. K. Oxenløw,  High-dimensional quantum key distribution based on multicore fiber using silicon photonic integrated circuits, npj Quantum Inf. {\bf 3}, 25 (2017)
\bibitem{Muller} A. Muller, T. Herzog, B. Huttner, W. Tittel, H. Zbinden, and N. Gisin, Plug and play systems for quantum cryptography, Appl. Phys. Lett. {\bf 70}, 793 (1997)
\bibitem{Bedington} R.  Bedington, J.M. Arrazole and A. Ling, Progress in satellite quantum key distribution, npj Quantum Information {\bf 3} 30 (2017)
\bibitem{Cozzolino} D. Cozzolino, D. Bacco, B. Da Lio, K. Ingerslev, Y. Ding, K. Dalgaard,
P. Kristensen, M. Galili, K. Rottwitt, S. Ramachandran, and L. K. Oxenløwe, Orbital angular momentum states enabling fiber-based high-dimensional quantum communication, Phys. Rev. Appl. {\bf 11}, 064058 (2019)
\bibitem{Jin} D.Jin, Y. Guo, Y. Wang amd D. Huang, Parameter estimation of orbital angular momentum based continuous-variable quantum key distribution, J. Appl.Phys. {\bf 127}, 213102 (2020)
\bibitem{Canas} G. Ca\~nas, N. Vera, J. Cariñe, P. González, J. Cardenas, P. W. R. Connolly, A. Przysiezna, E. S. Gómez, M. Figueroa, G. Vallone, P. Villoresi, T. Ferreira da Silva, G. B. Xavier, and G. Lima,  High- dimensional decoy-state quantum key distribution over multicore telecommunication fibers, Phys. Rev. A 96, 22317 (2017)
\bibitem{Bethune} D. S. Bethune, W. P. Risk, Autocompensating Quantum Cryptography, New J. Phys. \textbf{4}, 42 (2002) 
\bibitem{Bal} D. Balado, J. Li\~nares, X. Prieto-Blanco, D.Barral, Phase and polarization autocompensating N-dimensional quantum cryptography in multicore optical fibers, JOSA B \textbf{36}, 2793  (2019) 
\bibitem{Bal2} D. Balado, J. Li\~nares, and X. Prieto-Blanco, Phase auto- compensating high-dimensional quantum cryptography in elliptical-core few-mode fibers, J. Mod. Opt. {\bf 66}, 947 (2019)
\bibitem{Hwang} W. Hwang,  Quantum Key Distribution with High Loss: Toward Global Secure Communication, Phys. Rev. Lett. \textbf{91}, 057901 (2003)
\bibitem{Ben-Aryeh}   Y. Ben-Aryeh and S. Serulnik, The quantum treatment of propagation in non-linear optical media by the use of temporal modes, Phys Lett A  {\bf 155}, 473 (1991) 
\bibitem{Yariv} D. M. Pepper,  D. Fekete and A. Yariv,  Observation of amplified phase-conjugate reflection and optical parametric oscillation by degenerate four wave mixing in  a transparent medium, Appl. Phys. Lett. \textbf{33}, 41 (1978)
\bibitem{LinaresNJP} J. Li\~nares, M. C. Nistal and D. Barral, Quantization of Coupled 1d Vector Modes In Integrated Photonics Waveguides, New J. Phys. {\bf 10}, 063023 (2008)
\bibitem{Mandel} L. Mandel  and  E. Wolf, {\em Optical Coherence and Quantum Optics} (Cambridge University Press, Cambridge,  1995)
\bibitem{teoremaSU(N)} M.Reck, A Zeilinger, H. Bernstein and P. Bertani, Experimental realization of any discrete unitary operator, Phys. Rev. Lett. {\bf 73}, 58  (1994) 
\bibitem{Soto} H. de Guise, O. Di Matteo and L. L. S\'anchez-Soto, Simple factorization of unitary transformations, Phys. Rev. A  {\bf 97}, 022328 (2018)
\bibitem{LinaresOL} J. Li\~nares, X. Prieto-Blanco, C. Montero-Orille and V. Moreno,  Spatial mode multiplexing/demultiplexing by Gouy phase interferometry, Opt. Lett. {\bf 42}, 93 (2017)
\bibitem{OAMsorter}  J. Leach, M.J. Padgett, S. M. Barnett, S. Franke-Arnold, and\,J.Courtial, Measuring the orbital angular momentum of a single photon, Phys. Rev. Lett. {\bf 88}, 257901 (2002)
\bibitem{Berry} M. V. Berry, Interpreting the anholonomy of coiled light, Nature {\bf  326}, 266  (1987)



\end{thebibliography}
\end{document}